\ifdefined\XeTeXversion\else\pdfoutput=1\fi % arXiv: force pdfLaTeX for the JPEG photos (guarded so local XeTeX/tectonic still builds)
\documentclass{IEEEcsmag-preprint}

\usepackage[colorlinks,urlcolor=blue,linkcolor=blue,citecolor=blue]{hyperref}
\expandafter\def\expandafter\UrlBreaks\expandafter{\UrlBreaks\do\/\do\*\do\-\do\~\do\'\do\"\do\-}
\usepackage{upmath}
\usepackage{amssymb}
\usepackage{tabularx}
\usepackage{pgfplots}
\pgfplotsset{compat=1.18}
\usetikzlibrary{positioning,arrows.meta,fit,backgrounds,decorations.pathreplacing}

\definecolor{accent}{HTML}{00629B}     % IEEE Blue (Pantone 3015C): bars, lines, fills
\definecolor{accentdk}{HTML}{003366}   % IEEE dark navy: borders and label text
\definecolor{ieeeorange}{HTML}{FF6600} % IEEE Orange (accent): escape / key figure
\definecolor{muted}{HTML}{C2CDD9}      % light gray-blue: de-emphasized categories

\newcolumntype{Y}{>{\raggedright\arraybackslash}X}

\newcommand{\bilalphoto}{\includegraphics[width=1in,height=1.25in,clip,keepaspectratio]{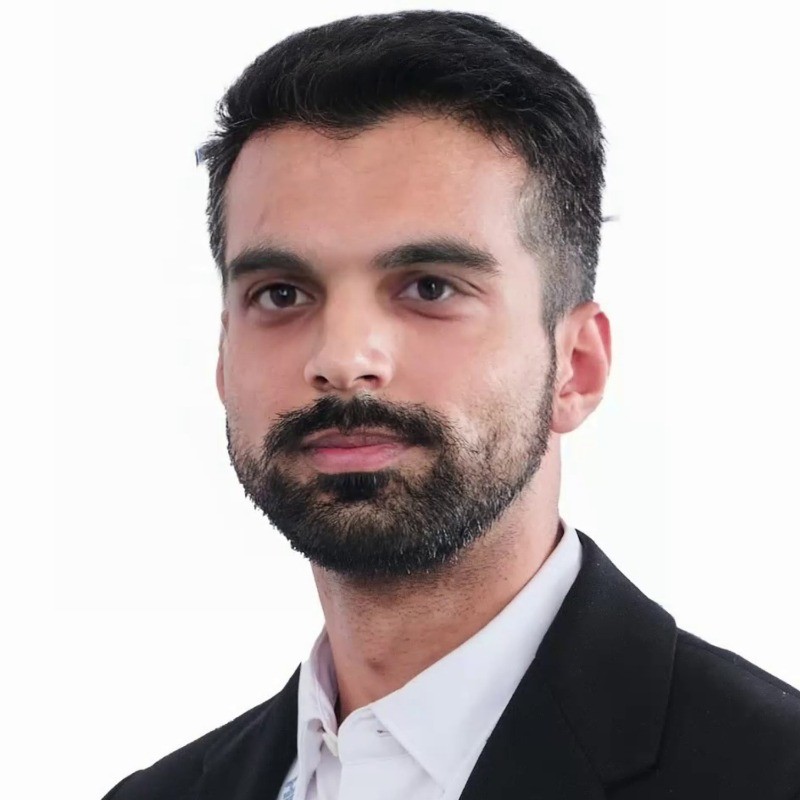}}
\newcommand{\aliphoto}{\includegraphics[width=1in,height=1.25in,clip,keepaspectratio]{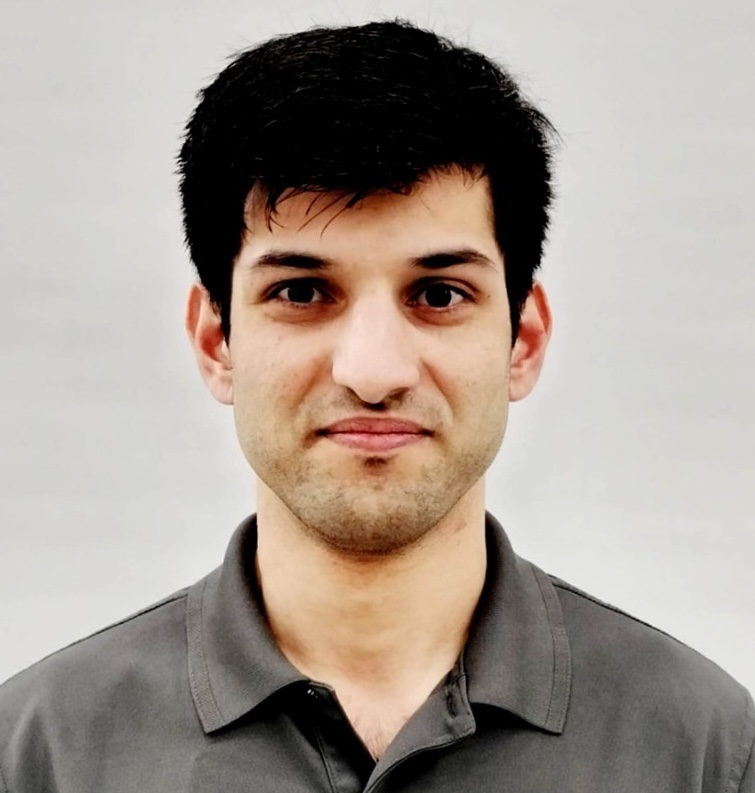}}

\jname{IEEE Software}
\jtitle{IEEE Software}
\jmonth{}
\pubyear{2026}
\jvol{}
\jnum{}
\paper{1}

\begin{document}

\sptitle{Feature Article: Software Testing}

\title{All Green, Still Broken: Real-Flow Verification Lessons from an
LLM-Integrated, Multi-Market Web Application}

\author{Muhammad Bilal}
\affil{Technical University of Munich, Germany}

\author{Ali Hassaan Mughal}
\affil{Independent Researcher, USA}

\markboth{FEATURE ARTICLE}{ALL GREEN, STILL BROKEN}

\begin{abstract}
Modern web applications increasingly combine three ingredients that are hard to
test: output from large language models, multi-market internationalization, and
browser-driven front-ends over external data sources. We report on a
production rental-search assistant whose automated suite grew to 1,553 test cases
in six weeks. The suite passed continuously, yet user-facing defects continued to
reach production. We studied all 252 bug-fix commits in the project and classified
each by the boundary, or seam, it escaped through. About 44 percent of the fixes
fall in four seams that component-level unit tests cannot observe: the live browser
runtime, the non-default market, the end-to-end flow, and the whole-system level.
A fix without a guard at the seam let one defect ship twice. We present the four-seam framework, the
measured defect distribution, and the practices we adopted, including a simple way
for a team to find the seam that carries the most fixes.
\end{abstract}

\maketitle

\section{KEY INSIGHTS}
\begin{itemize}
\item[{\ieeeguilsinglright}] A large, passing test suite can miss whole classes of
defect. A component test replaces the uncontrolled side of a boundary with a
stand-in to run deterministically. That stand-in hides the seam. In the system studied,
about 44 percent of bug fixes landed in such seams, so test count is a weak measure of
quality.
\item[{\ieeeguilsinglright}] A single end-to-end pass through the running product,
performed as a user in a non-default market, observes seams that component-level tests
cannot.
\item[{\ieeeguilsinglright}] When the same defect recurs, fix the whole class behind it. Add a build check that fails
on any repeat, so the same defect cannot ship again.
\end{itemize}

\chapteri{T}he system studied here is a production web application developed over a
six-week period. Its automated test suite grew to 1,553 test cases across 144 files
and was run on every change, with a passing run required before deployment.

During the same period the project recorded 252 bug-fix commits. Six
user-facing defects reached production. One recurred
after an earlier fix. All six passed the existing suite before release.

This article examines why a large passing suite did not prevent these defects. We
classified all 252 bug-fix commits in the project. Most escaped defects sat at boundaries
the components do not control: the browser runtime, the non-default market, and the
end-to-end user flow. We call these boundaries seams. We then measure how poorly
component-level tests observe them.

Prior work has already shown that coverage correlates weakly with how many defects
a suite catches.~\cite{inozemtseva2014} This article makes three contributions:
i)~a four-seam framework for the under-studied setting of one system that combines model
output, multi-market internationalization, and a browser runtime; ii)~a full-census
classification of 252 defects by the seam through which each escaped; and iii)~a
reproducible procedure that identifies the seam carrying the most fixes in a repository. Figure~1 summarizes
the argument.

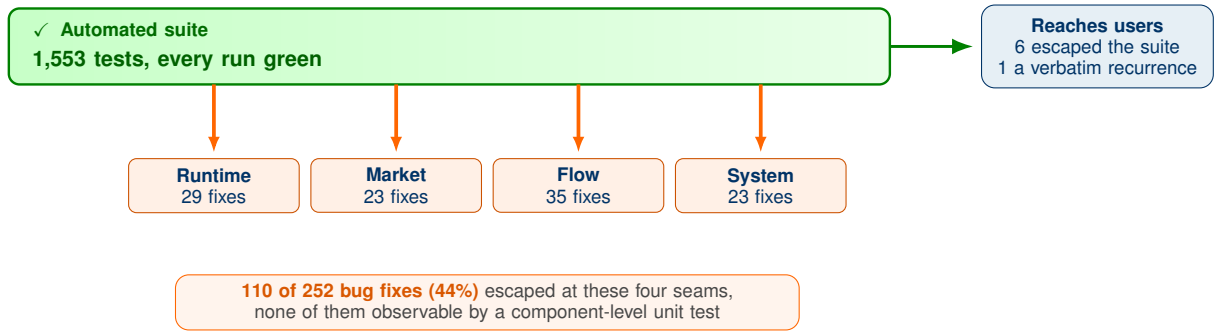
\begin{figure*}[t]
\centerline{\resizebox{\textwidth}{!}{%
\begin{tikzpicture}[
  font=\footnotesize\sffamily,
  leak/.style={-{Latex[length=6pt]}, ieeeorange, line width=1.4pt},
  chip/.style={draw=ieeeorange!80!black, fill=ieeeorange!10, rounded corners=3pt,
    align=center, inner sep=3.5pt, text width=2.0cm, font=\scriptsize\sffamily,
    text=accentdk},
]
% --- headline motif ---
\node[anchor=west, font=\large\sffamily\bfseries, text=accentdk] at (0,2.75) {All green, still broken};
\node[anchor=west, font=\scriptsize\sffamily, text=black!58] at (0,2.22)
  {A continuously passing suite leaks defects at four boundaries it never observes.};

% --- the "green" pipe: the passing suite ---
\shade[left color=green!22, right color=green!9, rounded corners=4pt] (0,0.55) rectangle (11.6,1.55);
\draw[draw=green!50!black, line width=0.9pt, rounded corners=4pt] (0,0.55) rectangle (11.6,1.55);
\node[anchor=west, font=\scriptsize\sffamily\bfseries, text=green!42!black] at (0.2,1.28)
  {\checkmark\, Automated suite};
\node[anchor=west, font=\footnotesize\sffamily\bfseries, text=green!38!black] at (0.2,0.86)
  {1{,}553 tests, every run green};

% --- flow onward to users ---
\draw[-{Latex[length=8pt]}, green!50!black, line width=1.1pt] (11.6,1.05) -- (12.7,1.05);

% --- four leaks at the seams (census counts) ---
\foreach \x/\name/\cnt in {2.7/Runtime/29, 5.1/Market/23, 7.5/Flow/35, 9.9/System/23}{
  \draw[leak] (\x,0.55) -- (\x,-0.35);
  \node[chip, anchor=north] at (\x,-0.42) {\textbf{\name}\\\cnt\ fixes};
}

% --- collecting tray: the 44% ---
\node[draw=ieeeorange, fill=ieeeorange!7, rounded corners=5pt, text width=8.0cm,
  align=center, font=\scriptsize\sffamily, text=black!72, anchor=north] at (6.3,-1.95)
  {{\color{ieeeorange!85!black}\bfseries 110 of 252 bug fixes (44\%)} escaped at these four
   seams, none of them observable by a component-level unit test};

% --- production / users ---
\node[draw=accentdk, fill=accent!10, rounded corners=5pt, align=center, inner sep=5pt,
  text width=2.7cm, font=\scriptsize\sffamily, text=accentdk, anchor=west] at (12.8,1.05)
  {\textbf{Reaches users}\\6 escaped the suite\\1 a verbatim recurrence};
\end{tikzpicture}}}
\caption{The paradox this article examines. An automated suite of 1,553 tests ran green on
every change, yet defects escaped at four boundaries that component-level tests never
observe. A full-census study of 252 bug-fix commits places 110 of them (44 percent) at
these four seams (Runtime, Market, Flow, and System), and six such defects
reached users, one of them a verbatim recurrence.}
\label{fig:overview}
\end{figure*}

\section{THE SYSTEM AND ITS TESTS}
The system is a rental-search assistant. It collects listings from several portals. It
screens each listing against a user's criteria with a large language model. A second
model drafts an outreach message. The results appear on a dashboard of server-rendered
HTML with small client-side libraries on top. The system serves two markets. The first
is the original market, with one European city, one currency, and one language. The
second was added later, with a different country, currency, and language.

Figure~2 shows how the suite grew. It reached 1,364 test functions, which pytest
expands to 1,553 test cases, in 43 days. The curve flattens after about three weeks.
All defects discussed in this report occurred after that point, while the suite was
already large and passing.

\begin{figure}
\centerline{%
\begin{tikzpicture}
\begin{axis}[
  width=\columnwidth, height=5.4cm,
  axis lines=left,
  xlabel={Days since first commit}, ylabel={Test functions (cumulative)},
  xmin=0, xmax=46, ymin=0, ymax=1700,
  xtick={0,7,14,21,28,35,42}, ytick={0,400,800,1200,1600},
  clip=false,
  ymajorgrids, grid style={gray!15}, tick align=outside,
  label style={font=\footnotesize\sffamily}, tick label style={font=\scriptsize\sffamily},
]
% region where every reported defect shipped (suite already large)
\addplot[draw=none, fill=ieeeorange!9, forget plot] coordinates {(21,0) (21,1700) (46,1700) (46,0)} \closedcycle;
% area under the growth curve
\addplot[draw=none, fill=accent!12, forget plot] coordinates {
(0,0) (2,128) (4,265) (7,387) (9,506) (12,681) (15,902) (18,1001)
(23,1022) (30,1057) (35,1118) (38,1211) (42,1315) (43,1364) (43,0)} \closedcycle;
% the growth curve
\addplot[accent, very thick, mark=*, mark size=1.0pt, mark options={fill=accent}] coordinates {
(0,0) (2,128) (4,265) (7,387) (9,506) (12,681) (15,902) (18,1001)
(23,1022) (30,1057) (35,1118) (38,1211) (42,1315) (43,1364)};
% plateau divider (marks ~week 3; named in the caption and the day-21 tick)
\draw[gray!55, dashed] (axis cs:21,0) -- (axis cs:21,1360);
% annotation placed in the empty upper-left triangle, never crossing the curve
\node[anchor=north west, align=left, font=\scriptsize\sffamily, text=ieeeorange!88!black, text width=3.4cm]
  at (axis cs:1.5,1640) {\textbf{All reported defects shipped after week 3,} when the suite was already large and green};
% final value callout, lifted clear above the end point
\node[anchor=south east, font=\scriptsize\sffamily\bfseries, text=accentdk] at (axis cs:43,1392) {1{,}364};
\end{axis}
\end{tikzpicture}}
\caption{Growth of the test suite over 43 days, counted as test functions. The suite
reached 1,364 functions, which pytest expands to 1,553 cases. Growth plateaus after
about three weeks; every defect in this report shipped in the shaded region that
follows, when the suite was already large and passing.}
\label{fig:growth}
\end{figure}
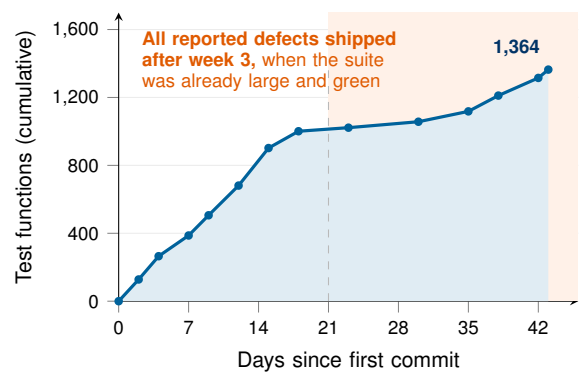

The suite is broad. About 30 percent of the tests cover data ingestion and parsing,
21 percent configuration and utilities, 10 percent operations, 10 percent
internationalization, 9 percent the dashboard, 8 percent the screening pipeline, 6
percent authentication, and 6 percent the drafter. The exercised code paths and the
actual product behaviors are not the same set. The defects below fall in that gap.

\section{THE FOUR SEAMS}
A seam is a boundary where our code meets something it does not control. We saw four
of them, shown in Figure~3.

The runtime seam is where server output meets a browser that runs scripts. Our tests
checked the HTML we sent. They did not run the browser that turns that HTML into
behavior.

The market seam is where default-market assumptions meet a user in another market.
Our tests and our manual checks used our own account, which always lived in the first
market.

The flow seam is where individually correct parts meet a real user interaction. A
handler and a template can each be correct on their own. The assembled interaction can
still fail.

The system seam is where a local change meets a whole-system rule. One component can
pass every test while a system-level guard misreads the new behavior.

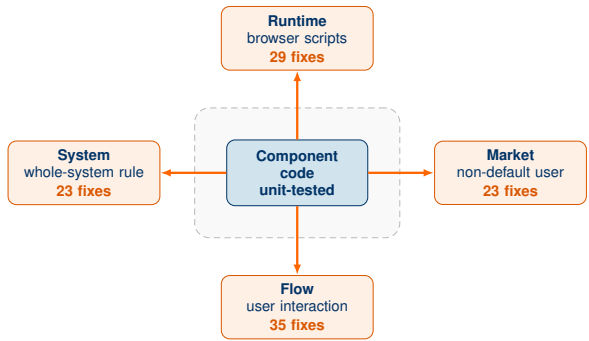
\begin{figure}
\centerline{\resizebox{\columnwidth}{!}{%
\begin{tikzpicture}[
  font=\scriptsize\sffamily, >={Latex[length=5pt]},
  core/.style={draw=accentdk, fill=accent!18, rounded corners=4pt, align=center,
    inner sep=5pt, text width=2.0cm, font=\scriptsize\sffamily\bfseries, text=accentdk},
  seam/.style={draw=ieeeorange!85!black, fill=ieeeorange!10, rounded corners=4pt,
    align=center, inner sep=4pt, text width=2.25cm, text=accentdk},
  esc/.style={-{Latex[length=5pt]}, ieeeorange, line width=1.1pt},
]
\node[core] (core) {Component code\\unit-tested};
% the region a component-level test actually observes
\begin{scope}[on background layer]
  \node[draw=gray!55, densely dashed, rounded corners=7pt, fit=(core), inner sep=15pt,
    fill=gray!5] (bnd) {};
\end{scope}
\node[seam, above=1.15cm of core] (rt)
  {\textbf{Runtime}\\browser scripts\\[1pt]{\color{ieeeorange!85!black}\bfseries 29 fixes}};
\node[seam, below=1.15cm of core] (fl)
  {\textbf{Flow}\\user interaction\\[1pt]{\color{ieeeorange!85!black}\bfseries 35 fixes}};
\node[seam, left=1.1cm of core] (sy)
  {\textbf{System}\\whole-system rule\\[1pt]{\color{ieeeorange!85!black}\bfseries 23 fixes}};
\node[seam, right=1.1cm of core] (mk)
  {\textbf{Market}\\non-default user\\[1pt]{\color{ieeeorange!85!black}\bfseries 23 fixes}};
\draw[esc] (core) -- (rt);
\draw[esc] (core) -- (fl);
\draw[esc] (core) -- (sy);
\draw[esc] (core) -- (mk);
\end{tikzpicture}}}
\caption{The shaded dashed region shows what a component-level unit test observes: one
component, with the rest stubbed out. The four seams lie just outside it, where the
code meets conditions it does not control, and the orange arrows show defects escaping
across that boundary. Counts are bug fixes per seam from the census of Figure~4.}
\label{fig:seams}
\end{figure}

The four seams share one mechanism. A component test swaps the
uncontrolled side of each boundary for a stand-in: a stored HTML string for a live
browser, the default configuration for the market axis, a stub for an adjacent component,
a fixed constant for the whole-system baseline. The swap makes the test repeatable. It
also removes the seam from view. A seam is the slice of behavior a test gives up to gain
determinism. Adding more such tests buys confidence inside the stubbed world. It does not
move the boundary of what gets observed. That is why the suite stayed large and green
while every defect in this report shipped.

\section{A STUDY OF 252 DEFECTS}
We classified every bug-fix commit in the project against the four seams.

\subsection{Method}
The classification followed four steps. i)~Sampling. We took one snapshot of the
repository: 740 commits across six weeks. From these we selected every commit whose
message begins with the conventional-commit prefix ``fix.'' That gave 252 bug-fix commits
against 272 feature commits, nearly one fix per feature. ii)~Taxonomy derivation. We
defined the four seams from the six initial defects (Table 1), then applied them unchanged to
the full census. iii)~Classification. A short list of priority-ordered keyword rules
assigned each fix to one category. A fix reaches a seam only when its message uses that
seam's vocabulary; everything else falls into a general backend group. The rules are
deterministic and contain no project-specific names.\footnote{Replication package (classifier rules and aggregate counts) archived at
Zenodo: \url{https://doi.org/10.5281/zenodo.20780007}.} We did not
hand-edit the output. A deterministic classifier has no inter-rater variance, so the
threat here is rule validity, not reliability. We treat the four-seam share as a
conservative lower bound, for the reasons in Study Limitations. iv)~Guard proxy. We
recorded whether each fix also changed a test file, as a proxy for adding a regression
guard.

\subsection{Results}
Figure~4 shows the distribution. About 110 fixes, or 44 percent, fall in the four
seams that unit tests cannot see: 35 in the flow seam, 29 in the runtime seam, 23 in
the market seam, and 23 in the system seam. Data-source fixes are 17 percent (44 fixes). The largest single group, 39 percent, is
ordinary backend logic, where component tests are strong. No single seam dominates.
Together the four seams total 110 fixes, more than double the data-source churn that
teams already treat as routine work.

\begin{figure}
\centerline{%
\begin{tikzpicture}
\begin{axis}[
  width=\columnwidth, height=5.8cm,
  xbar, xmin=0, xmax=120, axis lines=left,
  xlabel={Bug-fix commits (of 252)},
  symbolic y coords={System,Market,Runtime,Flow,Data,Backend},
  ytick={System,Market,Runtime,Flow,Data,Backend},
  nodes near coords, nodes near coords style={font=\scriptsize\sffamily\bfseries,
    text=accentdk, /pgf/number format/precision=0},
  tick label style={font=\scriptsize\sffamily}, label style={font=\footnotesize\sffamily},
  enlarge y limits=0.16, bar width=11pt, xmajorgrids, grid style={gray!13},
]
% other categories (where component tests are strong) -- muted
\addplot[xbar, bar shift=0pt, fill=muted, draw=gray!60] coordinates {
(98,Backend) (44,Data)};
% the four seams -- highlighted
\addplot[xbar, bar shift=0pt, fill=ieeeorange, draw=ieeeorange!55!black] coordinates {
(35,Flow) (29,Runtime) (23,Market) (23,System)};
% brace grouping the four seams (endpoints extended past bar centers to enclose them)
\draw[decorate, decoration={brace, amplitude=5pt}, ieeeorange!85!black, line width=0.8pt]
  ([yshift=10pt]axis cs:42,Flow) -- ([yshift=-10pt]axis cs:42,System);
\node[anchor=west, align=left, font=\scriptsize\sffamily, text=ieeeorange!88!black]
  at ([yshift=11pt]axis cs:49,Market) {\textbf{110 fixes}\\(44\%, unit-\\invisible)};
\end{axis}
\end{tikzpicture}}
\caption{All 252 bug-fix commits classified by seam, using the generic published rules
and sorted by frequency. The four seam categories (Flow, Runtime, Market, System) sum
to 110 fixes, or 44 percent, none observable by a component-level unit test. ``Backend''
(ordinary server logic) and ``Data'' (external sites changing) are where component tests
are already strong.}
\label{fig:taxonomy}
\end{figure}
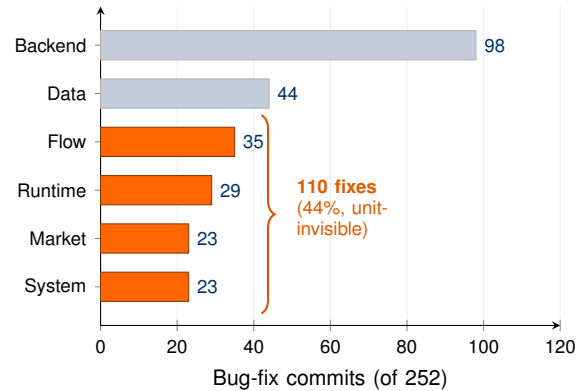

Of the 252 fixes, 107 (42.5 percent) also changed a test file. That is an upper bound on
real regression guards, since touching a test is not the same as adding the right one.
One defect still shipped twice. A quoting error in a client-side component broke a form
step. It was visible only in a running browser. The first fix left behind no test that
could catch it again. The guard lived in the same browser-blind harness that had missed
the defect, so it could not see the repeat. The 42.5 percent figure counts how often a
fix shipped with a test. It does not show whether that test sits where the defect can
return.

\subsection{Four Defects Up Close}
One defect from each seam follows.

In the runtime seam, a feature resolved a pasted listing link. It passed every
server-side test but failed in the browser. The search box sent its request dynamically,
and the handler replied with an ordinary redirect instead of the header the client
library expected. The page did not update.

In the market seam, a user in the second market tried to save a search and got an error.
A validator still encoded a rule from the first market and rejected a value that was
valid in the second. The branch had a passing unit test. No run had exercised it as a
user in the second market.

In the flow seam, the budget preset controls did not respond when selected. Each part
was correct in isolation. No test exercised the interaction.

In the system seam, ingestion volume was deliberately increased. A monitor compared
volume against a fixed baseline. It read the higher numbers as a broken pipeline and
raised a false alarm. Every test of the monitor passed, because each still assumed the
old baseline.

Table 1 lists all six, with at least one in each seam.

\begin{table*}
\caption{Six user-facing defects, grouped by the seam each exposed. All passed the
test suite before release; one recurred after an earlier fix.}
\label{tab:incidents}
\tablefont
\begin{tabularx}{\textwidth}{@{}c Y l Y@{}}
\toprule
\textbf{\#} & \textbf{What shipped broken} & \textbf{Seam} &
\textbf{Why the green suite missed it} \\
\colrule
1 & Pasted listing URL resolved server-side but failed over the dynamic
search box (it needed an HX-Redirect header, not a 303 redirect) & Runtime
& Validation never drove the real browser \\
2 & A real second-market user received HTTP 422 saving criteria; the
validator rejected a valid local listing type & Market & Verification walked only a
default-market user \\
3 & Onboarding overwrote a second-market user's chosen search type & Market
& Same blindness, a different surface \\
4 & A quoting error in a client-side component broke a form step from
initializing, a verbatim recurrence & Runtime & Visible only in a live browser \\
5 & Budget preset buttons were silently ignored in both markets & Flow &
Caught only by real interaction \\
6 & A false ``pipeline broken'' page appeared after we widened ingestion
volume & System & Unit tests did not model a system-level baseline shift \\
\botrule
\end{tabularx}
\end{table*}

\section{WHAT WE CHANGED}
We did not respond by writing more unit tests. We changed our definition of done and
where we spend test effort. We now treat an untested seam as technical debt to track and
pay down.~\cite{kruchten2012}

i)~Real-flow verification is now part of done. A change is finished only after we run the
real user action in the real runtime, across its states: a dynamic request and a plain
one, zero rows and many, each data source and an unsupported one, the success path and
each error path. In practice this is a short scripted pass through the running interface
in a browser.

ii)~We test along the market axis. Every market-aware screen is now exercised end to end
as a user in the non-default market. That configuration is the one our
routine development left uncovered.

iii)~We turn recurring defects into build checks. One guard
renders every screen for the second market and fails the build on any leak of the wrong
currency, the wrong city, or a first-language string. A companion check confirms that the
first market is unchanged. A refactor routed all market differences through one place, so
the guard has a single point to watch. The defect that had shipped twice can no longer
recur.

iv)~System guards must expect intentional change. We taught our throughput watchdog about
deliberate volume changes. A healthy change no longer trips an alarm meant for a broken
one.

Table 2 collects these habits as a checklist that other teams can apply. Each row
pairs a seam with what component tests miss and the lowest-cost check we adopted, and
points to the matching incidents in Table 1.

\begin{table*}
\caption{Seam-coverage checklist. Each row is backed by a practice we adopted and the
incidents that prompted it (Table 1).}
\label{tab:checklist}
\tablefont
\begin{tabularx}{\textwidth}{@{}l Y Y c@{}}
\toprule
\textbf{Seam} & \textbf{What component tests miss} & \textbf{The lowest-cost check we
adopted} & \textbf{Example} \\
\colrule
Runtime / browser & Behavior that appears only when the browser runs scripts, such as
dynamic-request handling and client-side initialization & One scripted real-browser
pass for each changed screen, across its request and error states & Inc.\ 1, 4 \\
Market / i18n & Branches that a default-market user never reaches & Walk each
market-aware screen end to end as a user in a non-default market & Inc.\ 2, 3 \\
Flow / UX & The wiring between individually correct parts & Drive the real
end-to-end user interaction, not its parts in isolation & Inc.\ 5 \\
System / ops & Whole-system shifts caused by an intended local change & Make system
guards aware of intended changes and re-baseline on deliberate volume changes & Inc.\ 6 \\
Recurrence & A fixed defect returning because the guard was not at the seam & Promote
the defect class to a build-level check that fails on any repeat & Inc.\ 4 \\
\botrule
\end{tabularx}
\end{table*}

The checklist is not specific to this system. Run the classifier from the replication
package on a project's commit history. It reports which seam carries the most
fixes. That seam is where additional testing is most likely to reduce escaped defects.

\section{FACTORS SPECIFIC TO THIS SETTING}
Three properties of this class of system intensify the problem. Output from a language
model is hard to unit test. It is not deterministic, and its correctness is fuzzy, a
version of the test oracle problem.~\cite{barr2015} Developers compensate with many
narrow assertions,~\cite{swebench2024,llmse-survey,chen2025} even as the model keeps
changing.~\cite{azanza2025} Internationalization multiplies every flow by a market axis
that stays hidden unless someone walks it. Browser-driven front-ends move real logic into
a runtime the backend test harness never starts. Each property pushes effort toward
isolated tests and away from the seams. This project had all three.

\section{RELATED WORK}
The weak link between test coverage and bug finding is established. Inozemtseva and
Holmes showed that coverage is not strongly correlated with test-suite
effectiveness.~\cite{inozemtseva2014} That fits a suite that is large but blind to these
boundaries. A related limit is the test oracle problem: deciding whether a given output
is correct.~\cite{barr2015} The seam problem studied here is upstream of the oracle. At a
seam the test environment never produces the production condition. The input that would
expose the defect is never generated, and the oracle is never consulted. Coverage and
oracle quality both presuppose that the triggering condition arises; at a seam it does
not. The four seams map where that precondition fails in one modern stack.

A second line of work studies the link between a test and the behavior it protects.
Traceability methods recover which test exercises which code,~\cite{tctracer2022} and
regression analysis traces which change introduced a bug.~\cite{maesbermejo2024}
Curated fault datasets such as Defects4J~\cite{defects4j2014} and
BugsInPy~\cite{bugsinpy2020} pair each real bug with a triggering test. Assembling
those links by hand is costly. Our recurrence showed the same thing in the field: a fix
without a guard at the seam protects nothing. An untested seam is a form of technical
debt to track and pay down.~\cite{kruchten2012}

Internationalization testing has been studied mainly at the presentation layer. Empirical
work catalogs internationalization layout and string-resource defects in mobile
interfaces~\cite{escobar2020} and detects layout and configuration failures in web pages,
while explicitly setting aside non-presentation locale defects.~\cite{alameer2016} We
found no prior work that treats locale-dependent validation or business logic as a defect
seam. That is the market-seam class this report documents. The escape mechanism also
resembles variability bugs: defects that appear only under certain build configurations,
which configuration-oblivious analyses miss.~\cite{mordahl2019} The market seam is a
runtime-locale analogue.

\section{STUDY LIMITATIONS}
This is an experience report on a single application, developed and maintained by its
two authors. The incidents are self-reported. We count the
defects we found, so silent escapes are not measured. The seam classification is
automated from commit-message keywords, with no hand-editing. Classification from commit
messages is known to mislabel some commits,~\cite{herzig2013} so the result is
approximate. A fix is placed in a seam only when its message uses that seam's vocabulary.
Fixes worded differently fall into a general backend group, which holds 39 percent of
fixes. The four-seam share is therefore a conservative lower bound. Whether the four seams
generalize to other systems that combine model output, internationalization, and a
browser runtime remains untested.

\section{CONCLUSION}
Counting passing tests says little about product quality. What matters is
which seams between the code and the user have been exercised at all. A large share of
defects occur at these seams, where component tests cannot see them. A low-cost check
that does see them is a single end-to-end pass through the running product. Run it in the
configuration least represented during development, such as a non-default market.
The spread of defects across seams shows a team where to put its testing effort.

\def\refname{REFERENCES}

% NOTE: this class's IEEEbiography environment does not accept the photo
% argument correctly under this engine, so the photo bios are composed
% manually (photo at left, bold name + text at right) to match the magazine.
\vskip 16pt
{\fontsize{9}{12}\selectfont
\noindent
\begin{minipage}[t]{1in}\bilalphoto\end{minipage}\hspace{8pt}%
\begin{minipage}[t]{\dimexpr\columnwidth-1in-8pt\relax}\raggedright
\textbf{Muhammad Bilal} is an AI and Digitalization Consultant in the German industrial
sector. He holds a Master of Science in Management from the Technical University of
Munich, Germany, and has previously worked as a Software Engineer, Business Analyst, and
Product Owner. His research interests include the impact of technology on business
performance, product quality analytics, the automation of industrial pipelines, large
language models, and agentic AI systems. He is the corresponding author.
\end{minipage}
\par\vskip 12pt
\noindent
\begin{minipage}[t]{1in}\aliphoto\end{minipage}\hspace{8pt}%
\begin{minipage}[t]{\dimexpr\columnwidth-1in-8pt\relax}\raggedright
\textbf{Ali Hassaan Mughal} has worked as a Senior Software Developer and Team Lead at
Xpressdocs, and earlier at Paycom and a stealth robotics company. He is pursuing an
Applied MBA in Data Analytics at Texas Wesleyan University, USA, and holds an M.Sc. in
Computer Science from Kansas State University. His research interests include automated
software testing, web application quality assurance, applied machine learning, large
language models, and agentic AI systems.
\end{minipage}
\par}

\end{document}